\documentclass{article}

\usepackage{PRIMEarxiv}

\usepackage[utf8]{inputenc} 
\usepackage[T1]{fontenc}    
\usepackage{hyperref}       
\usepackage{url}            
\usepackage{booktabs}       
\usepackage{amsfonts}       
\usepackage{nicefrac}       
\usepackage{microtype}      
\usepackage{lipsum}
\usepackage{fancyhdr}       
\usepackage{graphicx}       
\usepackage{courier}

\usepackage{lipsum}
\usepackage{soul}
\usepackage{color}
\usepackage{float}
\usepackage{graphicx}
\usepackage{caption}
\usepackage{subcaption}
\usepackage{titlesec}
\usepackage{multirow}
\usepackage{amsmath}
\usepackage{soul,color}

\graphicspath{{figures/}}     

\usepackage{placeins}
\usepackage{soul,color}
\title{Designing novel protein structures using sequence generator and AlphaFold2
\thanks{\textit{\underline{Citation}}: 
\textbf{X. Agha, N.Fu, and J.Hu Protein design using sequence generator and AlaphFold2. DOI:000000/11111.}} 
}

\author{
   Xeerak Agha \\
  Department of Computer Science and Engineering\\
  University of South Carolina\\
  Columbia, SC, 29201, USA \\
  \And
   Nihang Fu \\
  Department of Computer Science and Engineering\\
  University of South Carolina\\
  Columbia, SC, 29201, USA \\
\And
 Jianjun Hu \thanks{Corresponding author: J.H. (http://www.cse.sc.edu/~jianjunh)}\\
  Department of Computer Science and Engineering\\
  University of South Carolina\\
  Columbia, SC, 29201, USA \\
  \texttt{jianjunh@cse.sc.edu} \\
}

\begin{document}
\maketitle

\begin{abstract}
Protein structures and functions are determined by a contiguous arrangement of amino acid sequences. Designing novel protein sequences and structures with desired geometry and functions is a complex task with large state spaces. Here we develop a novel protein design pipeline consisting of two deep learning algorithms, ProteinSolver and AlphaFold2. ProteinSolver is a deep graph neural network that generates amino acid sequences such that the forces between interacting amino acids are favorable and compatible with the fold while AlphaFold2 is a deep learning algorithm that predicts the protein structures from protein sequences. We present forty de novo designed binding sites of the PTP1B and P53 proteins with high precision, out of which thirty proteins are novel. Using ProteinSolver and AlphaFold2 in conjunction, we can trim the exploration of the large protein conformation space, thus expanding the ability to find novel and diverse de novo protein designs.

\end{abstract}

\section{Introduction}
Proteins perform a wide variety of functions inside cells such as DNA replication and catalyzing reactions. They drive some of the fundamental processes in our bodies such as the binding of oxygen to red blood cells when inhaled air reaches the lungs \cite{kuhlman2019advances}. Protein structures are amino acids arranged in a contiguous order to form different formations. These formations are known as protein folds and once arranged in multiple folds the final structure is a 3D protein structure. The function of the protein is defined by the structure of this protein, which is defined by the binding and molecular constraints in an amino acid chain such as some amino acids being hydrophilic (having an affinity to water), some being hydrophobic (having an aversion to water), and some only selectively binding to other amino acids. Hydrophobic amino acids are found towards the inside of the protein structure formation to create a separation between the water molecules and the overall protein structure. Understanding how proteins fold from a given amino acid sequence can greatly help design novel proteins with desired structures and functions \cite{jumper2021highly,cao2022design}.

Protein design is a very complex problem in protein bioinformatics with a very large scope and search space, and has wide applications in the pharmaceutical industry, medicine, genetic engineering \cite{khoury2014protein}, synthetic biology\cite{gainza2018computational,zhou2020synthetic}, enzyme design for industrial process engineering \cite{broom2020ensemble}. Novel proteins are needed to produce drugs, vaccines, and enzymes for diseases such as diabetes, cancer, and COVID-19, amongst other industrial products \cite{mccarthy2013genomic, jacobson2004comparative, strokach2019designing}. There are several categories of algorithms for protein design as reviewed recently \cite{marcos2018essentials,setiawan2018recent,pan2021recent,frappier2021data}. These include first principles based de novo design, algorithms based on assembling short backbone fragments, and template based design. Recently, the data driven computational protein design \cite{frappier2021data,ferguson2021100th,siedhoff2021pypef} and especially the deep learning based protein design methods \cite{pearce2021deep,pan2021recent,ferruz2022controllable,cao2022design,courbet2022computational} emerged as one of the major progresses in this area. 


Current researches on structural bioinformatics have made significant accomplishments in the past two years with the capability to predict the protein structure from an amino acid sequence (protein sequence) with highly accurate results as done by the AlphaFold \cite{jumper2021highly} and RoseTTAfold \cite{baek2021accurate}, partially solving the so-called protein folding problem. However, designing a protein sequence from a target protein structure, which is called the inverse protein folding/design problem, has yet to have a high-accuracy solution \cite{pan2021recent}. Deep learning approaches such as using NLP transformers \cite{ferruz2022towards} and using different neural network backbones \cite{huang2022backbone} are being studied for the inverse protein folding problem. The pharmaceutical industry primarily uses in vitro experiments (not inside the body, but in similar conditions) to determine the sequence and function of newly designed proteins \cite{strokach2019designing}. However, such experiments are labor-intensive and expensive, and do not allow the testing of a wider variety of proteins. A challenge for the computational design of protein structures using such experiments is the large size of the protein conformation space \cite{meconi2022key}. In some methods, such as the Monte Carlo (MC) method, the encoding of the force fields in the simulation also increases the size of conformation space making it hard to determine protein folding. For example, the Monte Carlo identifies the lowest energy state of the protein chain, which is hard for large protein chains, thus increasing the size of the protein folding space. In addition, a structural prediction is only as accurate as the physics and structural embedding of the protein structure. Thus, exacerbating the size of the already huge protein folding state space, which has yielded a huge demand for Machine Learning/Artificial Intelligence (ML/AI) solutions for protein design problems. 

The key step in the protein design problem is to determine a link between amino acid sequences and protein structures, and one way to do that is to determine the possible amino acid sequences that a structure can yield. Researchers aim to explore faster and more accurate machine learning models to predict protein sequences that fold into a target structure. Studies have used the Markov-Chain Monte Carlo method to generate a protein sequence optimized to a force-field or statistical potential \cite{strokach2019designing}. However, the accuracy of these models has remained low. ProteinSolver \cite{strokach2020fast} is a Deep Graph Neural Network (DGNN) model for generating novel protein sequences that can later conform to protein structures. The DGNN uses a weighted distance matrix to determine the distance between amino acids and their interactions. ProteinSolver can identify the amino acids with respect to the local geometric and chemical environment in a protein template. The model tests amino acids' positions in many proteins and then uses that data to predict amino acid sequences masked over a protein structure template. However, ProteinSolver does not have the capability to generate the structures. The main goal of this study is to use the ProteinSolver for the protein sequence generation and then use Alphafold2 for the structure generation. We apply our method to design the local binding sites for the Protein Tyrosine Phosphatase (PTP1B) protein and tumor suppressor (P53) protein to facilitate drug-target binding. Here, PTP1B is one of the main protein targets for type II diabetes treatment. Type II diabetes drugs that target PTP1B reduce cell body insulin resistance and lead to an uptake of insulin in the cells from the bloodstream in type II diabetic patients \cite{he2014protein}. And P53 protein is the main protein that prevents cancer. The loss of the P53 tumor suppressor pathway leads to cancer because it is responsible for programmed cell death (Apoptosis), DNA repair, and cell cycle arrest \cite{shangary2009small}. Drugs targeting the P53 protein regulate the P53 protein and thus have a chance to restore regular cell growth and function in the human body. The deep graph neural network (DGNN) generates novel sequences of the PTP1B and P53 Protein PDB structure input. The algorithm produces feasible sequences in a large search space of protein conformations. Our study uses a unique protein design pipeline with two modules, ProteinSolver and AlphaFold2, to generate novel protein sequences for PTP1B and P53 proteins that can later be folded into structures to be used in therapeutic drugs for type II diabetes and cancer respectively. However, it is important to note that if given wrong sequences or masking parameters, our pipeline is unlikely to produce the desired sequences. Our study is successful in generating thirty novel protein sequences out of the forty total protein sequences generated for both PTP1B and P53 proteins. 


\section{Related Works}
This work is related to deep learning based protein design. The first related work is template-based protein design, which is one of the main categories of algorithms for protein design. Here, backbone designability is mainly governed by side chain-independent interactions showing the need for designing new backbones based on continuous sampling and optimization of the backbone-centered energy surface. New backbones are built either by parametrically varying relative geometries between existing structural modules (or templates) to design helix bundles or repeat proteins, or by assembling peptide fragments from existing structures. Despite recent improvements \cite{jacobs2016design, pan2020expanding}, the template dependence of these approaches for generating backbones still severely restricts the available spectrum of possible new structures, potentially narrowing the scope of the functional activities amenable to design. Huang et al. \cite{huang2022backbone} developed a statistical model, SCUBA (Side Chain-Unknown Backbone Arrangement) that uses neural network-form energy terms. SCUBA outlines design constraints such as conformational preferences, hydrogen-bonding geometry of peptide backbones and inter-backbone space, and tightly packed side chains with high precision. High accuracy is achieved through a two-step process: first, the statistical energy values are estimated from raw structural data using kernel-based density estimation, and secondly, the neural network (fully connected three-layer perceptron) is trained to represent the potentials. SCUBA can display high-dimensional and highly correlated distributions of real structural data with high fidelity. SCUBA allows continuously sampled and optimized structures with complete flexibility. A key innovation of SCUBA is that it trims the sequence search space in the backbone design stage. Thus, SCUBA allows a comprehensive search of novel backbone architectures that are not observed in nature. 

With the progress of protein structure prediction enabled by deep neural networks, new protein design methods have been developed to take advantage of these tools. Anishchenko et al. \cite{anishchenko2021novo} explored how to use neural network-based methods to generate new folded proteins with sequences that are unrelated to the naturally occurring protein used to train the model. A random amino acid sequence is generated and put into the trRosetta structure prediction network to predict the starting residue (residue distance maps), which is featureless in the beginning. Monte Carlo sampling is then done on the amino acid sequence space and the contrast is optimized using the Kullback-Leibler (KL) Divergence test between the inter-residue distance distributions predicted by the network and background distributions averaged over all proteins. Random starting points of the neural network allows a wide range of sequences and predicted structures. The results obtained are gene encodings of 129 network-'hallucinated' sequences. The proteins are  expressed and purified in Escherichia coli (E. coli) and 27 of the proteins resulted in consistent similarities with the hallucinated proteins. Two are tested using X-ray crystallography and one by Nuclear magnetic resonance spectroscopy (NMR) is also consistent with the hallucinated models. Thus, deep neural networks are train to predict protein structures from protein sequences can be inverted to design new proteins alongside physics-based models. 

Natural-language-based approaches are also being explored for protein design. Castro et al.  \cite{castro2022guided} developed an autoencoder, ReLSO, that allows explicit modeling of the sequence to function landscape within the large labeled datasets through optimization within the existing design space. Feruz and Höcker \cite{ferruz2022towards} developed natural language processing (NLP) based transformers for controllable protein sequence design. Just as letters form words and sentences carry meaning, amino acids can be arranged in a variety of combinations to form structures that carry function. Divide-and-conquer approaches where proteins are split into protein units (PU) and each PU is explored in a parallel fashion are also being explored \cite{pal2022modularity} for protein design. The increasing impact of deep learning in protein structure modeling is summarized by Baek and Baker \cite{baek2022deep}, which shows that deep learning tends to affect many areas of structural biology by further advances in it. Overall, we find that deep learning for computational protein design is becoming a staple in the field \cite{cao2022design, wang2018computational, pearce2021deep, ovchinnikov2021structure}. 

\section{Materials and Methods}

\subsection{ Protein design targets}

\begin{figure}[h] 
    \centering
    \includegraphics[width=0.60\textwidth, scale=.5]{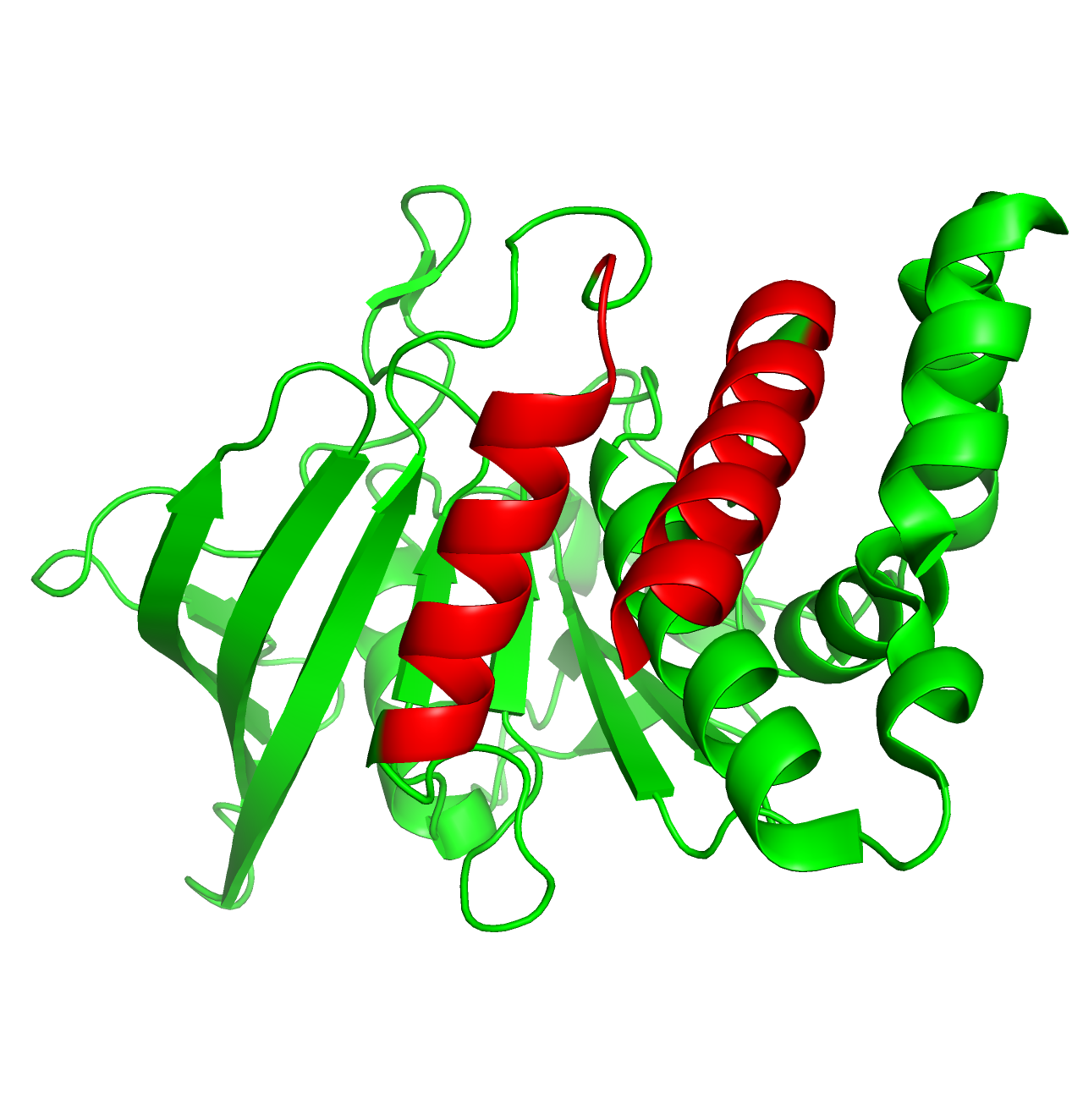}
    \caption{PTP1B protein structure. The masked locations are from 184-198 with the sequence of ESPASFLNFLFKVRE and the positions from 262-279 with the sequence of ADQLRFSYLAVIEGAKFI (shown in red).}
    \vspace{-3pt}
    \label{fig:ptp1b_allo}
\end{figure}

The protein design space is vast and dynamic. Here we focus on the local protein structure design: the binding site design problem. We choose to examine two proteins PTP1B and P53, which are common drug design targets with associated diseases. The PTP1B protein and P53 protein are associated with Type II Diabetes and Cancer respectively \cite{hussain2019protein}. PTP1B assists in the uptake of insulin and thus glucose in the body to treat the effects of diabetes. The P53 protein is responsible for controlling rapid cell growth that is exhibited in cancer and leads to the major effects of cancer. The PDB structure file of the Protein Tyrosine Phosphatase 1B (PTP1B) protein is retrieved from the Research Collaboratory for Structural Bioinformatics (RCSB) Protein Data Bank (PDB) \cite{bank1971protein, berman2000protein}. The RCSB PDB provides information about three-dimensional (3D) structures of macromolecules and associated small molecules in the PDB archive. The target structure and the binding site of PTP1B are shown in Figure\ref{fig:ptp1b_allo}.

As for the define the binding site design problem, we masked the \(\alpha\)3, \(\alpha\)6, and \(\alpha\)7 helices that comprise the allosteric binding site of the PTP1B target protein. The \(\alpha\)3 helix consists of amino acids Glu186-Glu200, the \(\alpha\)6 helix consists of amino acids Ala264-Ile281, and the \(\alpha\)7 consists of amino acids Val287-Ser295. The allosteric binding site is shown in Figure \ref{fig:ptp1b_allo}.

\begin{figure}[H]
    \centering
    \includegraphics[width=0.60\textwidth, scale=.5]{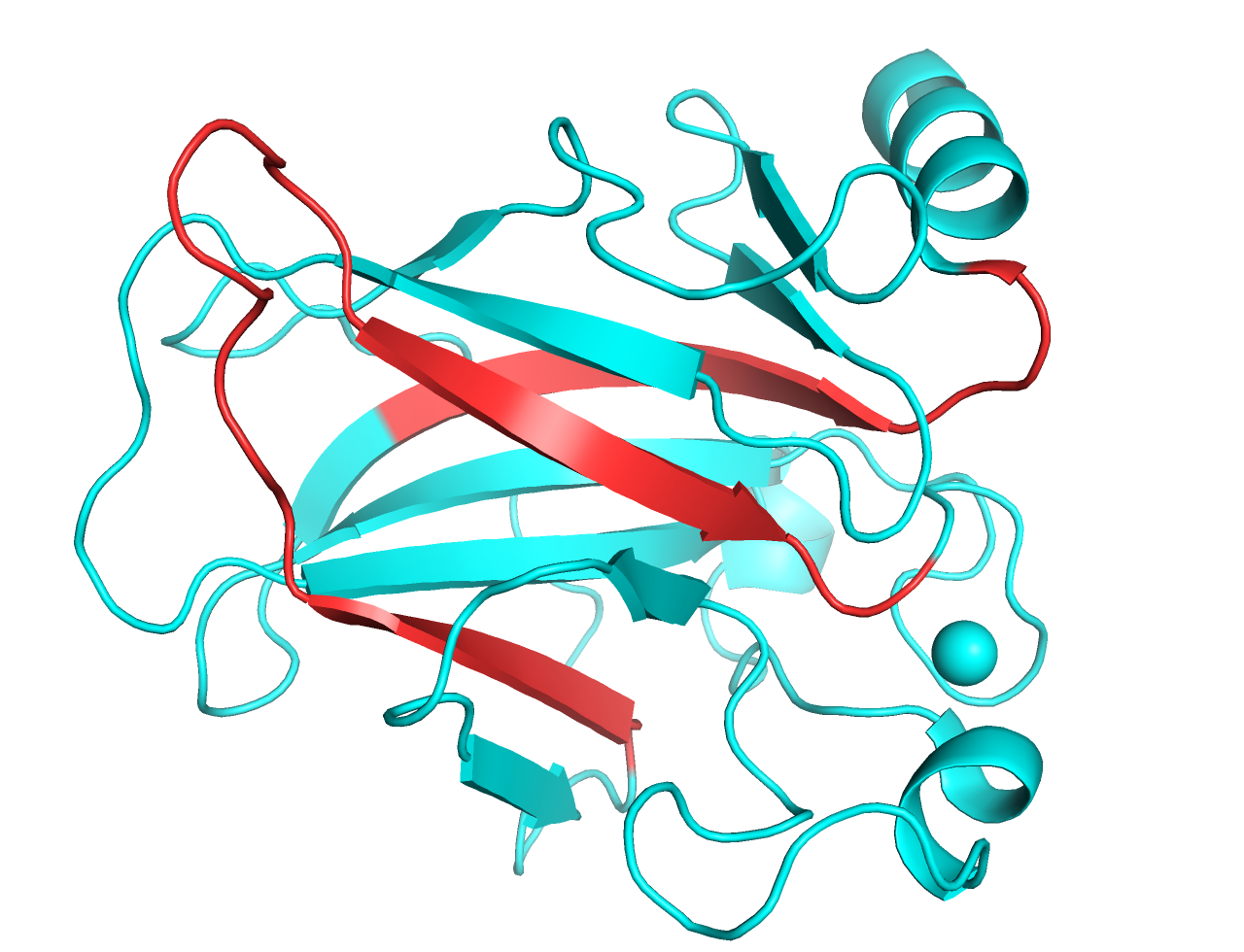}
    \caption{P53 protein structure. The masked locations are from 117-142 with the sequence of RHSVVVPYEPPEVGSDYTTIYFKFMC and the locations of 171-181 with the sequence RDSFEVRVCAC (shown in red).}
    \vspace{-3pt}
    \label{fig:p53_allo}
\end{figure}

For the P53 tumor suppressor protein, we first retrieve its target PDB structure file from the RCSB PDB. We then masked the amino acids of the core binding domain that the peptide FL-CDB3 drug has bindings with. The core binding domain can be seen in Figure \ref{fig:p53_allo}. The core domain consists of amino acids Arg117-Cys142 and Arg171-Cys181. The design goal is to generate a binding domain that matches this target domain that we masked out.

\subsection{Design Pipeline}
 
\begin{figure}[h!] 
    \includegraphics[width=\textwidth]{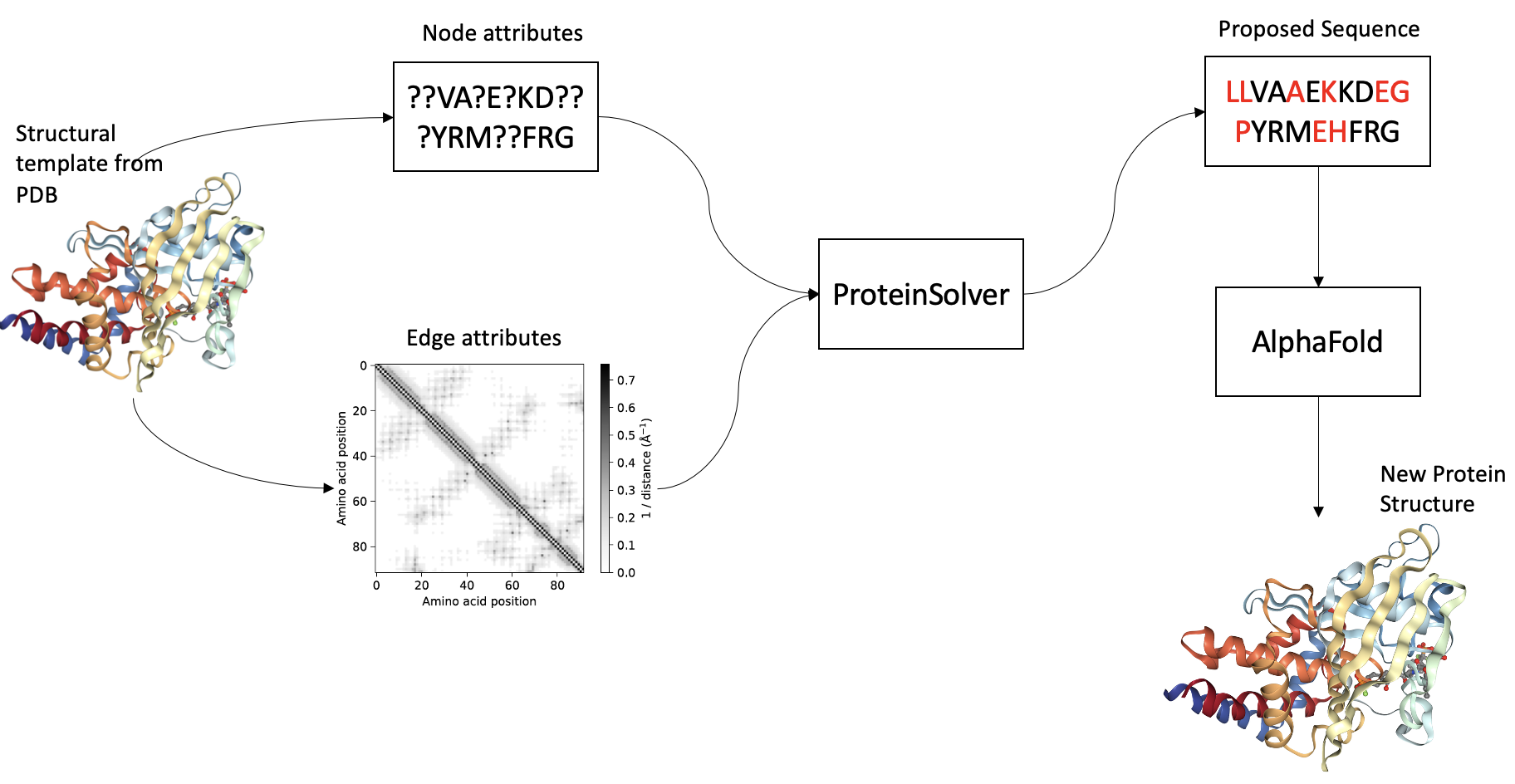}
    \vspace{-3pt}
    \caption{ Our protein design pipeline. It contains two major modules including ProteinSolver, which is used to generate novel protein sequences and AlphaFold, which is used to predict the structures of the generated protein sequences. ProteinSolver uses a deep graph neural network to generate novel protein sequences. AlphaFold2 uses neural network blocks called Evoformers that generate a folded 3D protein structure based on the protein sequence input. }
    \label{fig:pipeline}
\end{figure}

 Our protein design pipeline is composed of two modules including the ProteinSolver \cite{strokach2020fast} and the AlphaFold2 \cite{jumper2021highly} as shown in Figure\ref{fig:pipeline}. The ProteinSolver is a graph neural network classification model that can predict/generate the amino acid types at the masked positions of the template, which fit the best with the local environment. The Alphafold2 is a deep neural network model for predicting the protein structure given its sequence.  
 
 Our protein design pipeline takes a template protein structure as input and mask a specific local sequence of the allosteric binding site as the design target. We then generate a matrix to encoding the edge attributes of neighboring amino acids, then feed the masked sequence with a given length and the edge attribute matrix to the the graph neural network of the ProteinSolver to generate a variety of allosteric binding site sequences for the PTP1B target protein. We run the ProteinSolver algorithm to generate 20 candidate amino acid sequences that satisfy the spatial constraints within the template for the PTP1B target protein. We then run these 20 sequences through the AlphaFold2 algorithm \cite{jumper2021highly}  to generate the corresponding structures for these sequences. The structural integrity of the structures is verified using AlphaFold2. The root means square deviation (RMSD) is calculated between the generated structure and the original target structure using superposition in PyMOL, which is an open-source molecular visualization tool \cite{yuan2017using}. Superposition is conducted with the original protein structure and the newly generated protein structure by AlphaFold2. Once superimposed, the difference between each atom of the original and newly generated proteins is calculated in the Cartesian space to calculate an overall RMSD value. In our design problem here, the binding site of the PTP1B therapeutic target protein is masked so that we can compare the original binding site structure to the generated binding site structure by their RMSD to evaluate the protein design quality of our pipeline. For the P53 core binding domain, we apply the same procedure to generate candidate design structures. 

 The novelty of Our protein design pipeline is that, without the deep graph neural network (DGNN) of the ProteinSolver, the space complexity of finding a novel amino acid sequence for the binding site of the PTP1B protein would be $20^{39}$, where 39 is the number of amino acids open for design. The DGNN algorithm finds the solution in a linear space complexity of O(N) where N is the length of the sequence to be designed, instead of the exponential space complexity of $20^{N}$. The full design pipeline can be seen in Figure \ref{fig:pipeline}. 

\subsection{ProteinSolver architecture and implementation}
\begin{figure}[h!] 
    \includegraphics[width=\textwidth]{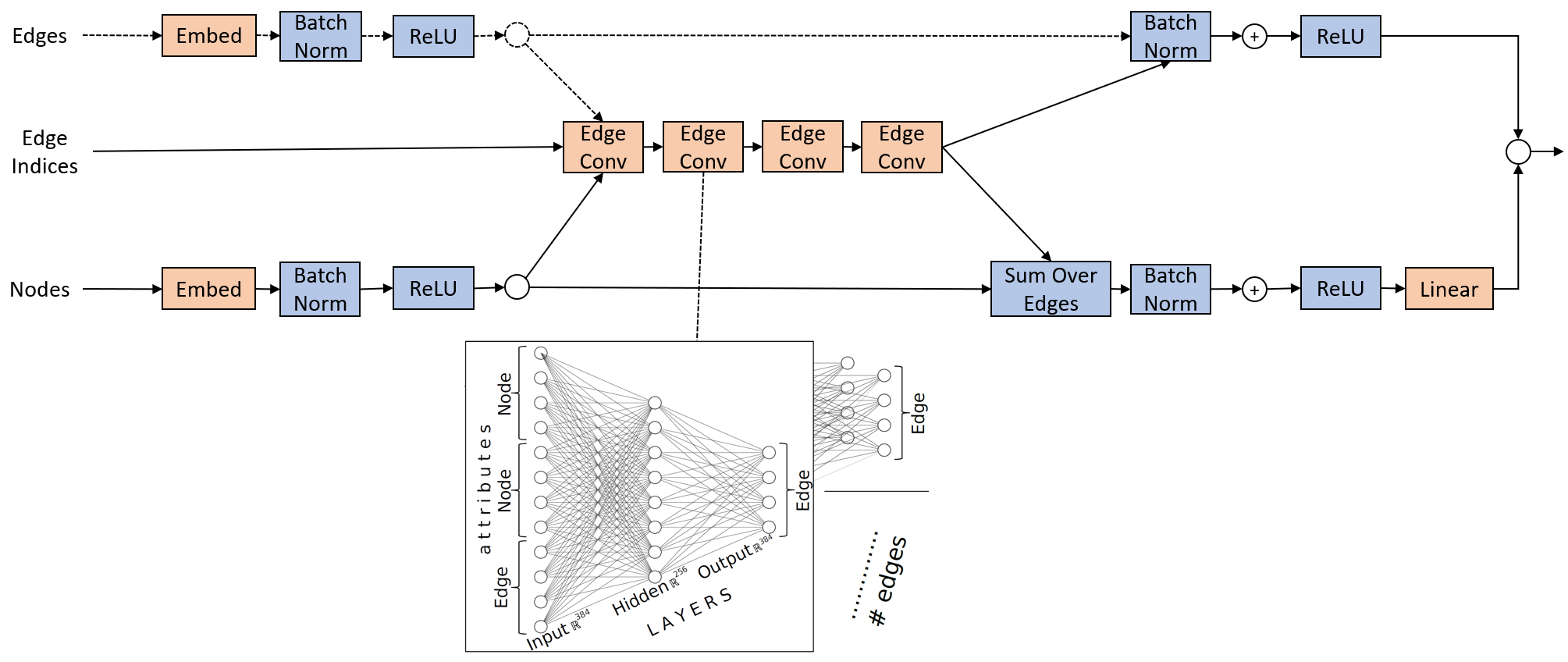}
    \vspace{-3pt}
    \caption{ProteinSolver architecture. The ProteinSolver architecture takes in the edge and node attributes of the protein structure separately. It then applies an embedding layer, batch normalization, and Rectified Linear Unit (ReLU) on both the edges and nodes. Next, the edge, nodes, and edge indices go through a series of edge convolution (EdgeConv) blocks. The edges then go through batch normalization and ReLU once more whereas the nodes go through a sum of edges layers, batch normalization, ReLU, and a linearization layer. Finally, the edge and node embeddings are the results. The edge embeddings are where each amino acid link is in the overall protein sequence. The Node embeddings go through ReLU and a softmax function to finally get the node label (amino acids).}
    \label{fig:NN_Arch}
\end{figure}

\FloatBarrier
ProteinSolver is a graph neural network based generation model, which can determine the candidate amino acid sequences within a protein 3D structure as shown in Figure \ref{fig:NN_Arch}. ProteinSolver can design the protein sequences that fold into a predetermined shape by phrasing this challenge as a constraint satisfaction problem (CSP), which is similar to Sudoku puzzles \cite{strokach2020fast}. 
The inputs to the ProteinSolver network are a set of node attributes and a set of edge attributes that can describe intersections between pairs of nodes. The nodes and edges are embedded using a multi-layer perceptron in m-dimensional space, then they are passed through several Edge Convolution (EdgeConv) blocks that produce new node and edge attributes with the same dimensions as the inputs but they now have information about the immediate neighbors of each node. The new node and edge attributes are added to the inputs from the last layer and the resulting tensors are passed through a ReLU layer to update the input node and edge attributes. After ReLU the tensors are passed into a subsequent graph convolution residual block, with four residual blocks being applied. Here, the most optimal number of edge convolution layers is 4. Finally, the node attributes are passed through ReLU nonlinearity and into a final Linear layer. This maps 162 features describing each node in N-dimensional space, where N is the number of outcomes which in the case of protein design is 20. The outputs of the Linear layer can optionally be passed through a softmax function to convert raw outputs values into probabilities of a given node being a given specific label. The function of the EdgeConv blocks is to concatenate the edge attributes and attributes of the interacting nodes and passes the resulting vector through a multi-layer perceptron. The outputs of the perceptron from the new edge attributes and the summation of the edge attributes of each node produce new node attributes. The node and edge attributes are normalized using the BatchNorm layer and are added to the input tensors which complete the residual blocks. 

The architecture and hyperparameters of the ProteinSolver network are optimized by training the network to solve Sudoku puzzles formulated from the constraint satisfaction problem. This formulation is used because it is a well-defined problem for which predictions made by the network can be easily verified \cite{yuan2021systematic, hu2019strategies, hu2020pretraining}. The graph neural network is trained to reconstruct missing numbers in the Sudoku grid while minimizing the cross-entropy loss between predicted and actual values. After optimizing the architecture for solving Sudoku puzzles the same network is applied to the protein design. Proteins are treated as graphs where nodes are individual amino acids and edges are the shortest distances between pairs of amino acids considering that the pairs of amino acids are 12\AA of one another. The node attributes identify amino acid types as well as a flag to indicate if the amino acid type is not known. The edge attributes include the shortest distance between each pair of amino acids in Cartesian space and the number of residues separating the pair of amino acids along the protein chain. The model is trained with approximately half of the amino acids in the sequence missing and the network is trained to reconstruct the masked amino acids by minimizing the cross-entropy loss between predicted and actual values. The accuracy of the network in reconstructing amino acid sequences from training and validation datasets, 50\% of the amino acids are marked missing, is tracked and the training process is stopped once the validation accuracy plateau. The results of the training show that despite 80\% of the original residues being given as input, the generated sequences share little homology with the original residues.

\subsection{AlphaFold2}
AlphaFold2 is a computational model that predicts protein structures with atomic accuracy even when a similar structure is not present. AlphaFold2 uses novel neural network architectures and training procedures based on evolutionary, physical, and geometric constraints of protein structures. AlphaFold2 uses domain-level knowledge in combination with a novel deep neural network architecture to generate protein structures with the high accuracy \cite{alquraishi2019alphafold}. The AlphaFold2 network is comprised of two stages. The first stage is the novel neural network block (Evoformers) processing and the second stage is the iterative refinement of 3D shape propositions. In the first stage the base of the network takes the inputs through repeated layers of a novel neural network block that DeepMind has termed Evoformers \cite{jumper2021highly, beattie2016deepmind}. The Evoformer produces $N_{seq}$ x $N_{res}$ array ($N_{seq}$ = number of sequences and $N_{res}$ = number of residues) pairs. There is an array representation for the multiple sequence alignment (MSA) and residues. Evoformer blocks contain attention-based and non-attention-based components. The key benefit of using Evoformer blocks is that they contain mechanisms to exchange information within the MSA and pair representations, which enables direct reasoning about spatial and evolutionary relationships. In the second stage after the trunk of the network, there is a structural module that shows the rotation and translations of each protein residue (global rigid body frames) introducing an explicit 3D protein structure. The rotations are then initialized to a trivial state where all rotations are set to the identity and all positions are set to the origin. The global rigid body frames rapidly develop into a refined and highly accurate protein structure. The benefit of this structural module is that the chain structure can be broken to allow simultaneous local refinement of any parts of the protein structure. An equivalent transformer allows the neural network to reason about the underrepresented side-chain atoms and loss term that places weight on the correctness of the orientation of the residues. 

AlphaFold2 is used as a highly accurate protein structure predictor for globular proteins of industrial and biomedical interest \cite{pinheiro2021alphafold}. AlphaFold2 has also been used for multi-chain protein complex prediction by training the AlphaFold2 on multi-chain folded structures \cite{evans2021protein}. Multimeric inputs of the known stoichiometry increase the accuracy of the predicted multimeric interfaces while maintaining high intra-chain accuracy. For 17 heterodimer proteins without templates, AlphaFold2 achieves at least medium accuracy (continuous protein-protein docking model quality score (DockQ $\geq$ 0.49). For 13 heterodimer proteins without templates, AlphaFold2 achieves high accuracy (DockQ $\geq$ 0.8). As for heteromeric interfaces AlphaFold2 can predict interfaces (DockQ $\geq$ 0.23) in 70\% of cases to produce high accuracy predictions (DockQ $\geq$ 0.8) in 26\% cases \cite{evans2021protein}. AlphaFold2 has the capability of defining protein folding for multiple single-chain and multi-chain proteins. Thus, AlphaFold2 is heavily used in the field of protein design and computational biology as a whole due to its robustness. 

\section{Results and Discussion}
We started with a masked template protein structure for the protein structures to generate a target candidate structure. Once the target candidate structure is generated we used RMSD to measure the performance of our pipeline. The root means squared deviation$^{1}$ (RMSD) of the AlphaFold2 generated structures is calculated for each generation in comparison to the original protein structure using superposition in PyMol. The RMSD value greater than one means that the protein structure designed is not the same as the original protein structure, whereas the RMSD value less than one means the designed protein structure is closer to the original protein structure. For example, in Table \ref{tab:alkali}, except for the generated samples 2 and 12 of PTP1B and the generated samples 0, 5, 15, and 19 of P53 which have the RMSD smaller than one, most other generations show the RMSD value greater than one, which means that there are a lot of novel generations. The average root means squared deviation (aRMSD) for the PTP1B protein is 1.433 \AA\space, and the P53 protein is 1.606 \AA. These results show that on average our design pipeline is able to generate new protein sequences that can later be synthesized into protein structures for real-world applications. 

\begin{equation}
rmsd = \sqrt{(\frac{1}{T})\sum_{t=1}^{T}(y_{t} - \hat{y_{t}})^{2}}
\label{eq:rmsd}
\end{equation}

RMSD is calculated using Eq.\ref{eq:rmsd}, where $T$ which is the number of non-missing data points, $t$ which is the variable being used in the summation function, $y_{t}$ which is the actual observation, and $\hat{y_{t}}$ which is the estimated observation. This difference between the estimated observation and actual observation is taken in Cartesian space using Cartesian distances. The sum of those Cartesian distances is the RMSD value. The summary of the RMSD values for the therapeutic target proteins can be seen in Table \ref{tab:alkali}.


The study has produced several new protein structures. The generated structures are shown in Table \ref{tab:alkali}. Two designed PTP1B candidate and four designed P53 candidate protein structures were highly similar to the target structures. Site 1 and Site 2 were highly similar in the generated candidate 2 and 12 protein sequences for the PTP1B target and generated candidate 0, 5, 15, and 19 protein sequences for the P53 target protein. PTP1B generated candidate 2 had 4 base pair matches in site1 and 11 base pair matches in site2 to the PTP1B target shown in Table \ref{tab:alkali}. P53 generated candidate 0 had 10 base pair matches in site1 and 4 base pair matches in site2 to the P53 target as shown in Table \ref{tab:alkali}. Seventeen PTP1B and fourteen P53 protein structures were slightly different to the target structures. PTP1B generated candidate 4 has 6 base pair matches in site 1 and 10 base pair matches in site 2 to the PTP1B target shown in Table \ref{tab:alkali}. P53 generated candidate 11 has 7 base pair matches in site 1 and 3 base pair matches in site2 to the P53 target as shown in Table \ref{tab:alkali}. One PTP1B and two P53 protein structures were extremely different to the target structures. PTP1B generated candidate 14 had 5 base pair matches in site1 and 7 base pair matches in site2 to the PTP1B target as shown in \ref{tab:alkali}. P53 generated candidate 2 had 6 base pair matches in site1 and 3 base pair matches in site2 to the P53 target as shown in \ref{tab:alkali}.

\begin{table}[hbt!]
\centering
\caption{The root means squared deviation (RMSD) value for the novel therapeutic target protein generations}
\begin{tabular}{ |c|c|c|c|} 
 \hline
 Target Protein and generated & site1 & site2 & RMSD (\AA)   \\
 \hline \hline 
 \textbf{PTP1B target} & \textcolor{blue}{\texttt{ESPASFLNFLFKVRE}} & \textcolor{blue}{\texttt{ADQLRFSYLAVIEGAKFI}} & 0 \\
 generated 2 & \texttt{LDLRE\hl{F}VQ\hl{FL}WL\hl{V}WR} & \texttt{PS\hl{QLRFSY}Q\hl{AV}LD\hl{GA}RN\hl{I}} & 0.979 \\
 generated 12 & \texttt{DN\hl{PA}AH\hl{L}F\hl{FL}KQIWK} & \texttt{PKRFK\hl{F}C\hl{Y}R\hl{AV}L\hl{EG}G\hl{K}E\hl{I}} & 0.981 \\
 generated 3 & \texttt{DGGDD\hl{F}IK\hl{F}VWS\hl{V}YR} & \texttt{PA\hl{QLRFSY}E\hl{A}L\hl{I}T\hl{GAK}E\hl{I}} & 1.016 \\
 generated 10 & \texttt{TG\hl{P}TNMME\hl{FL}RECY\hl{E}} & \texttt{PQKM\hl{RF}A\hl{Y}E\hl{A}L\hl{I}S\hl{G}S\hl{K}EV} & 1.049 \\
 generated 13 & \texttt{DN\hl{P}KA\hl{F}ID\hl{F}ILS\hl{V}Y\hl{E}} & \texttt{PN\hl{QL}H\hl{F}A\hl{Y}Q\hl{AVI}C\hl{GAK}QV} & 1.050 \\
 generated 9 & \texttt{TT\hl{P}SG\hl{FL}DMMQLVWN} & \texttt{PTEF\hl{RFSY}E\hl{A}IVAAS\hl{K}H\hl{I}} & 1.155 \\
 generated 19 & \texttt{RD\hl{P}ED\hl{FL}A\hl{F}IM\hl{KV}YK} & \texttt{\hl{A}SE\hl{LRF}A\hl{YLAV}LS\hl{G}S\hl{K}IL} & 1.169 \\
 generated 15 & \texttt{IN\hl{P}KN\hl{F}CE\hl{F}MFQ\hl{V}W\hl{E}} & \texttt{PA\hl{Q}F\hl{RF}AWNCL\hl{L}VA\hl{A}ER\hl{I}} & 1.170 \\
 generated 7 & \texttt{KN\hl{P}SA\hl{F}FL\hl{F}IQR\hl{V}WL} & \texttt{PSE\hl{L}K\hl{F}T\hl{YL}GC\hl{IEGAK}KL} & 1.193 \\
 generated 8 & \texttt{AD\hl{P}FD\hl{FL}E\hl{FLFKV}YQ} & \texttt{PEEF\hl{RF}A\hl{Y}I\hl{AV}LV\hl{G}VEKL} & 1.197 \\
 generated 0 & \texttt{TN\hl{PA}D\hl{FL}D\hl{FL}K\hl{KV}Y\hl{E}} & \texttt{NCEMK\hl{F}C\hl{Y}KA\hl{V}MK\hl{GA}RL\hl{I}} & 1.250 \\ 
 generated 11 & \texttt{ED\hl{P}QE\hl{F}VK\hl{FL}Y\hl{KV}W\hl{E}} & \texttt{PQEFK\hl{F}C\hl{Y}M\hl{A}ILA\hl{GA}SE\hl{I}} & 1.253 \\
 generated 18 & \texttt{DD\hl{P}EK\hl{F}IK\hl{FLF}S\hl{V}YA} & \texttt{PE\hl{QL}K\hl{FSY}I\hl{A}I\hl{IE}AY\hl{K}E\hl{I}} & 1.275 \\
 generated 1 & \texttt{NNGSA\hl{F}IN\hl{FL}RQ\hl{V}YD} & \texttt{PAE\hl{LRF}T\hl{Y}E\hl{AV}L\hl{E}A\hl{AK}R\hl{I}} & 1.276 \\
 generated 5 & \texttt{GD\hl{P}R\hl{SF}MD\hl{F}MY\hl{KV}YY} & \texttt{PA\hl{QL}K\hl{F}A\hl{Y}V\hl{AV}LHA\hl{AK}KL} & 1.317 \\
 generated 16 & \texttt{KN\hl{P}D\hl{SFL}K\hl{F}IT\hl{KV}Y\hl{E}} & \texttt{PN\hl{QLRF}T\hl{Y}E\hl{AV}LQA\hl{AK}DW} & 1.374 \\
 generated 17 & \texttt{DT\hl{P}NL\hl{FL}G\hl{F}SL\hl{KV}YK} & \texttt{ID\hl{Q}F\hl{R}YA\hl{Y}I\hl{AV}LKASRQV} & 1.448 \\ 
 generated 4 & \texttt{HT\hl{P}QE\hl{FL}Q\hl{F}IYH\hl{V}Y\hl{E}} & \texttt{TVEI\hl{RFSYLAVI}D\hl{GA}QKV} & 1.485 \\
 generated 6 & \texttt{AN\hl{P}Y\hl{SFL}E\hl{FLF}ACWT} & \texttt{GSE\hl{LRF}C\hl{Y}V\hl{A}LLHAS\hl{K}H\hl{I}} & 1.587 \\
 generated 14 & \texttt{DNAT\hl{SFL}E\hl{F}CMD\hl{V}WS} & \texttt{P\hl{DQL}K\hl{FSY}VSILQ\hl{G}CRFL} & 5.429 \\
 
 \textbf{P53 target} & \textcolor{blue}{\texttt{RHSVVVPYEPPEVGSDYTTIYFKFMC}} & \textcolor{blue}{\texttt{RDSFEVRVCAC}} & 0 \\
 generated 0 &\texttt{WM\hl{S}I\hl{VV}G\hl{YEPP}LA\hl{G}QPQ\hl{T}VLR\hl{F}RFTV} & \texttt{\hl{R}CTL\hl{EV}Y\hl{V}AQT} & 0.853 \\
 generated 5 & \texttt{WL\hl{SVV}TALQ\hl{PP}RH\hl{G}LSHS\hl{T}LR\hl{F}R\hl{F}IV} & \texttt{NC\hl{SFE}IHNSSK} & 0.889 \\
 generated 15 & \texttt{WFA\hl{V}T\hl{V}AFL\hl{PP}KQ\hl{G}ESQS\hl{T}LR\hl{F}L\hl{F}DT} & \texttt{QHC\hl{F}DT\hl{R}ITNL} & 0.903 \\
 generated 19 & \texttt{WRAC\hl{V}TAFR\hl{PP}RKGTGHDL\hl{I}R\hl{F}R\hl{F}DT} & \texttt{YSA\hl{FEV}HT\hl{C}ET} & 0.927 \\
 generated 8 & \texttt{WLA\hl{V}T\hl{V}GFS\hl{PP}IK\hl{G}QAHDYLR\hl{F}Y\hl{F}DV} & \texttt{HAG\hl{FE}TH\hl{V}TTD} & 1.015 \\
 generated 14 & \texttt{WM\hl{SV}EAGFL\hl{PP}RKDVGH\hl{T}V\hl{I}Q\hl{F}R\hl{F}EV} & \texttt{QSL\hl{FEV}H\hl{V}SGN} & 1.046 \\
 generated 18 & \texttt{YN\hl{SV}Y\hl{V}AFN\hl{PP}KS\hl{G}QSHQ\hl{T}LR\hl{F}IYTV} & \texttt{YTA\hl{F}D\hl{V}HSTDL} & 1.054 \\
 generated 6 & \texttt{W\hl{HSV}R\hl{VP}FY\hl{PPE}K\hl{G}VVNSILR\hl{F}QFI\hl{C}} & \texttt{\hl{R}RG\hl{F}I\hl{V}KTSTD} & 1.060 \\
 generated 17 & \texttt{WKV\hl{V}E\hl{VP}FH\hl{PP}LWNQAND\hl{T}L\hl{YF}M\hl{F}DT} & \texttt{HSTLN\hl{V}QTSSQ} & 1.085 \\
generated 4 & \texttt{WM\hl{SV}I\hl{V}GF\hl{EPP}LS\hl{G}KKIS\hl{T}LRY\hl{KF}LV} & \texttt{QL\hl{S}MD\hl{V}E\hl{V}AEE} & 1.169 \\
 generated 16 & \texttt{YF\hl{SV}R\hl{V}SFQ\hl{PP}VPARAED\hl{T}LR\hl{F}Y\hl{F}NT} & \texttt{ERA\hl{FEV}HTADS} & 1.170 \\
 generated 7 & \texttt{WLACT\hl{V}AFTE\hl{PE}DSIGHSI\hl{I}R\hl{F}N\hl{F}Y\hl{C}} & \texttt{H\hl{D}R\hl{F}D\hl{V}LTSGD} & 1.181 \\
 generated 10 & \texttt{WRG\hl{V}S\hl{V}GFD\hl{PP}LTNLSQSVLR\hl{FKF}NT} & \texttt{ENV\hl{FE}IKTA\hl{A}D} & 1.610 \\
 generated 11 & \texttt{WR\hl{SV}Q\hl{V}GFD\hl{PP}VDALLHNQLR\hl{F}I\hl{F}EV} & \texttt{EEM\hl{FEV}YTSDD} & 1.804 \\
 generated 12 & \texttt{FM\hl{S}IDCGFL\hl{PP}MNTI\hl{D}SNVVK\hl{F}F\hl{F}TT} & \texttt{QE\hl{SFEV}Y\hl{V}ASA} & 2.001 \\
 generated 9 & \texttt{WV\hl{SV}K\hl{VP}FS\hl{PP}YE\hl{GS}NRN\hl{T}IK\hl{F}TYDT} & \texttt{QTA\hl{FE}IY\hl{V}A\hl{A}D} & 2.089 \\
 generated 13 & \texttt{Y\hl{H}GIIT\hl{P}FQ\hl{P}IRE\hl{G}RPLDVVK\hl{F}T\hl{F}DT} & \texttt{\hl{R}MRV\hl{EV}ECTQD} & 2.115 \\
 generated 3 & \texttt{W\hl{HSV}A\hl{V}AFD\hl{P}A\hl{E}ET\hl{S}AQN\hl{T}LK\hl{F}N\hl{F}VS} & \texttt{QTA\hl{FEV}HTS\hl{A}D} & 2.384 \\
 generated 1 & \texttt{H\hl{HSV}STAFT\hl{PP}I\hl{V}EVACDQ\hl{I}R\hl{F}V\hl{F}DV} & \texttt{YY\hl{SFE}IHTSSD} & 3.865 \\
 generated 2 & \texttt{YSA\hl{V}R\hl{V}GFK\hl{P}ILKFE\hl{D}NDNVR\hl{F}I\hl{F}ET} & \texttt{EMT\hl{F}V\hl{V}Y\hl{V}STD} & 3.890 \\
 
 \hline
 \label{tab:alkali}
\end{tabular}
\end{table}

\newpage
\subsection{Generated Structures}
\begin{figure}[ht!] 
\centering
    \begin{subfigure}[t]{0.49\textwidth}
        \includegraphics[width=\textwidth]{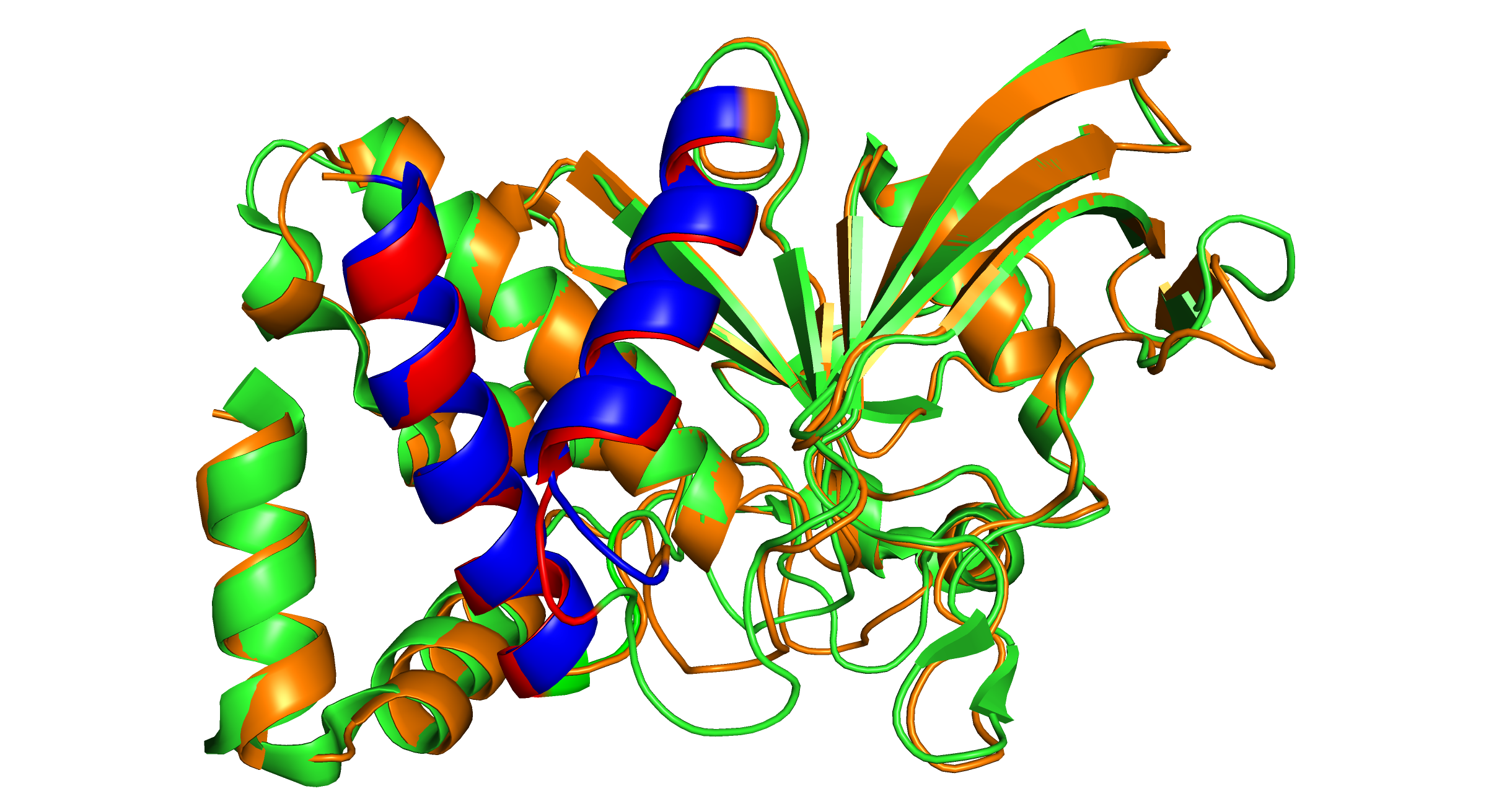}
        \caption{}
        \vspace{-3pt}
        \label{fig:hnov1}
    \end{subfigure}
    
   \caption{New protein designed based on PTP1B protein template. The 14th generation of the PTP1B protein has a RMSD of 5.429\AA \space as shown in Table \ref{tab:alkali}. This protein is novel as it has a RMSD far above 1.0\AA.}
  \label{fig:ptp1b_hnov}
\end{figure}

\begin{figure}[ht!] 
\centering
    \begin{subfigure}[t]{0.49\textwidth}
        \includegraphics[width=\textwidth]{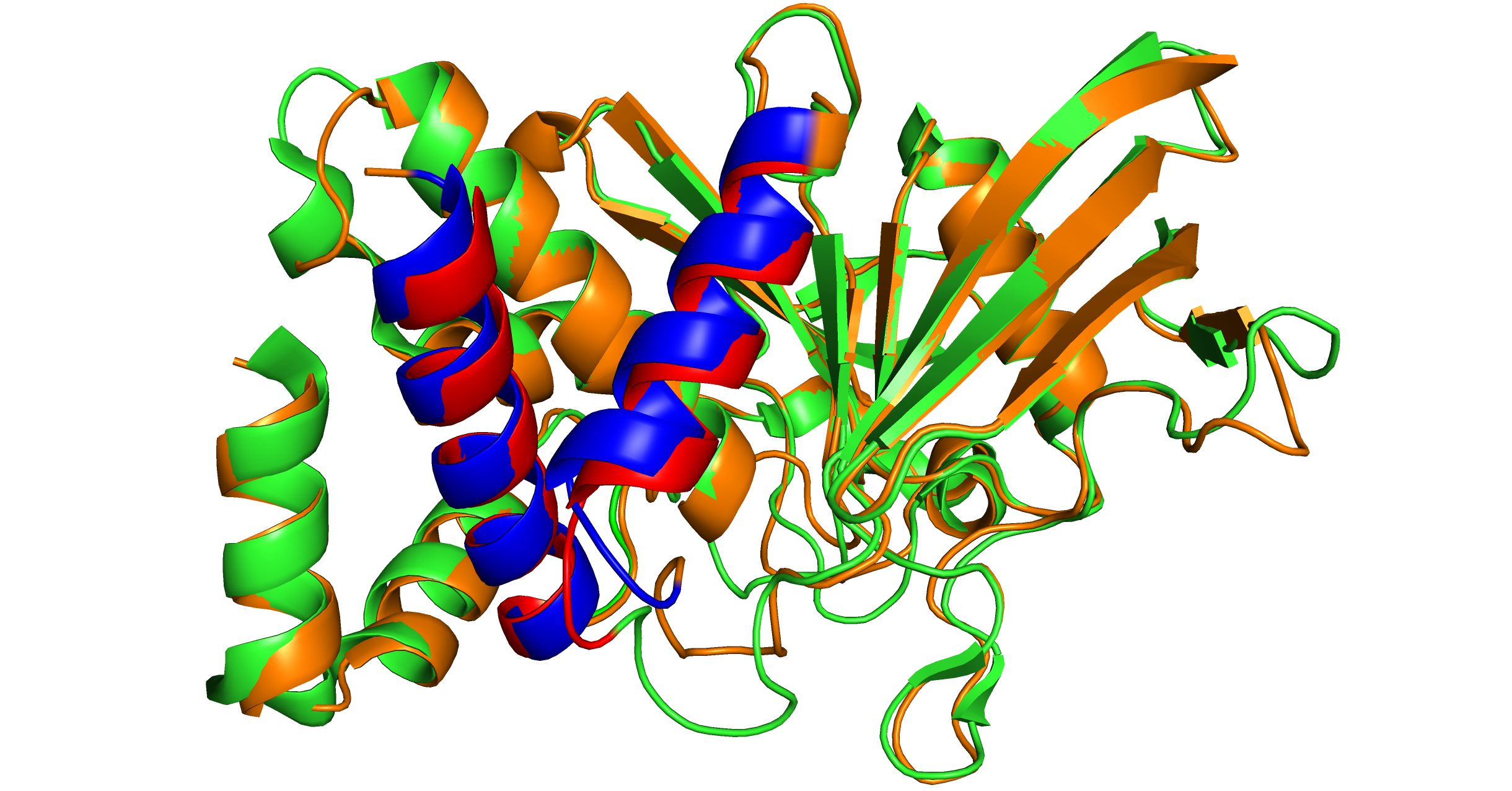}
        \caption{}
        \vspace{-3pt}
        \label{fig:nov1}
    \end{subfigure}
    \begin{subfigure}[t]{0.49\textwidth}
        \includegraphics[width=\textwidth]{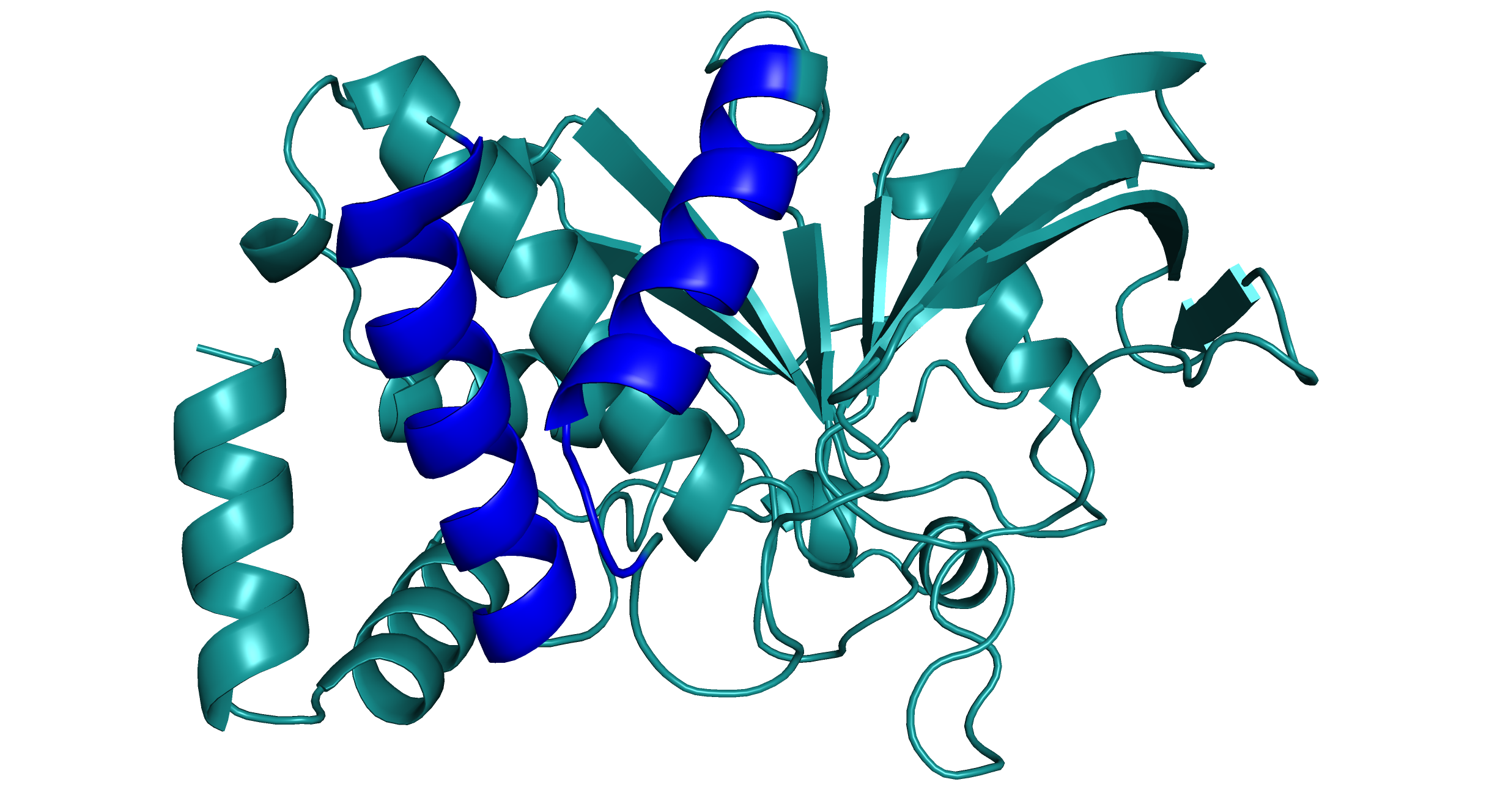}
        \caption{}
        \vspace{-3pt}
        \label{fig:nov2}
    \end{subfigure} 
    \begin{subfigure}[t]{0.49\textwidth}
        \includegraphics[width=\textwidth]{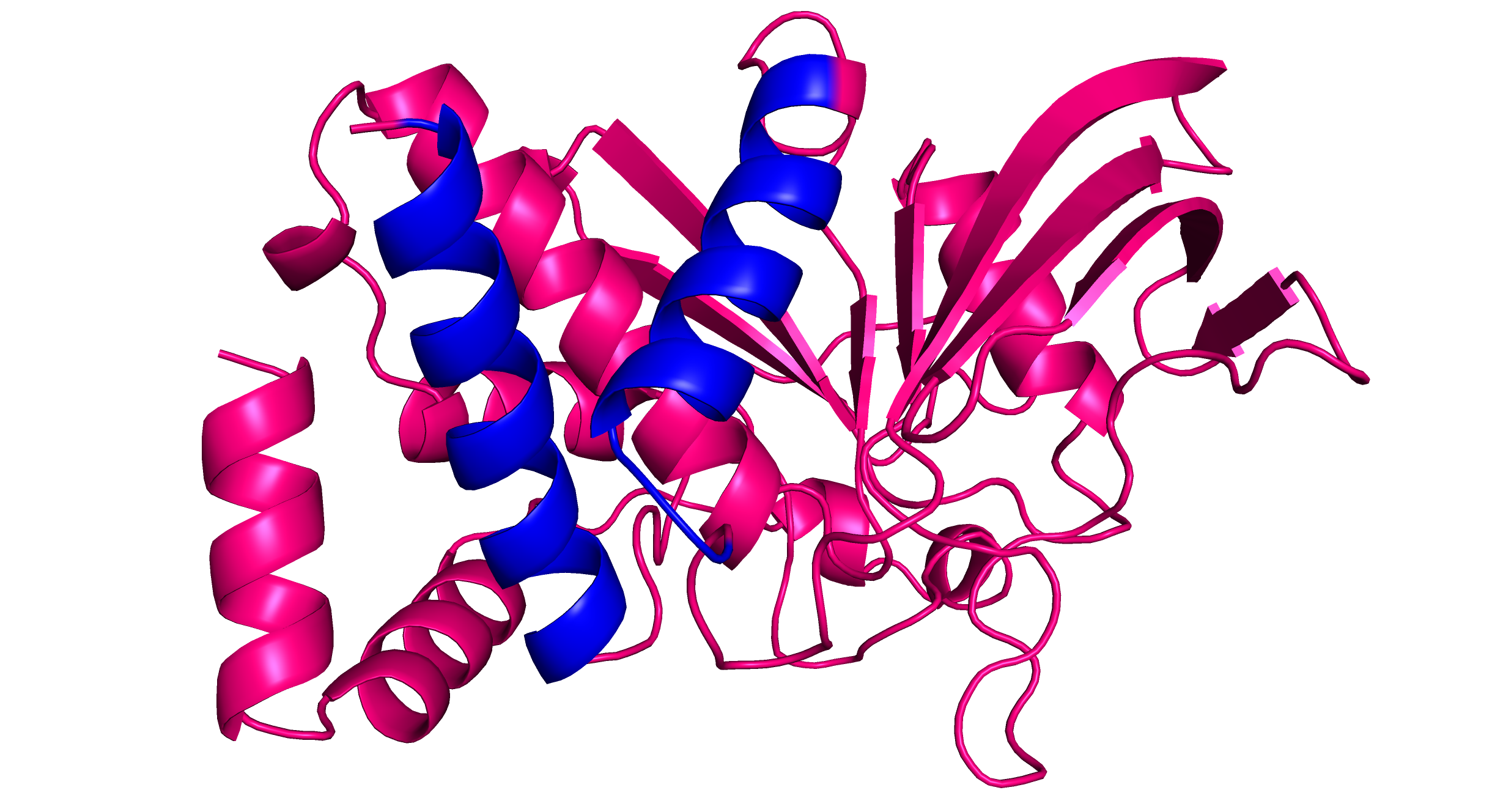}
        \caption{}
        \vspace{-3pt}
        \label{fig:nov3}
    \end{subfigure} 
    
   \caption{ New designed variants of the PTP1B protein. (a) The 4th generation of the PTP1B protein with RMSD 1.485\AA\space as shown in Table \ref{tab:alkali}. (b) The 6th generation of the PTP1B with RMSD 1.587\AA\space as shown in Table \ref{tab:alkali}. (c) The 17th generation of the PTP1B with RMSD 1.448\AA\space as shown in Table \ref{tab:alkali}. These generated proteins are novel because they have RMSD values above 1.0\AA.}
  \label{fig:ptp1b_nov}
\end{figure}

\begin{figure}[ht!] 
    \begin{subfigure}[t]{0.49\textwidth}
        \includegraphics[width=\textwidth]{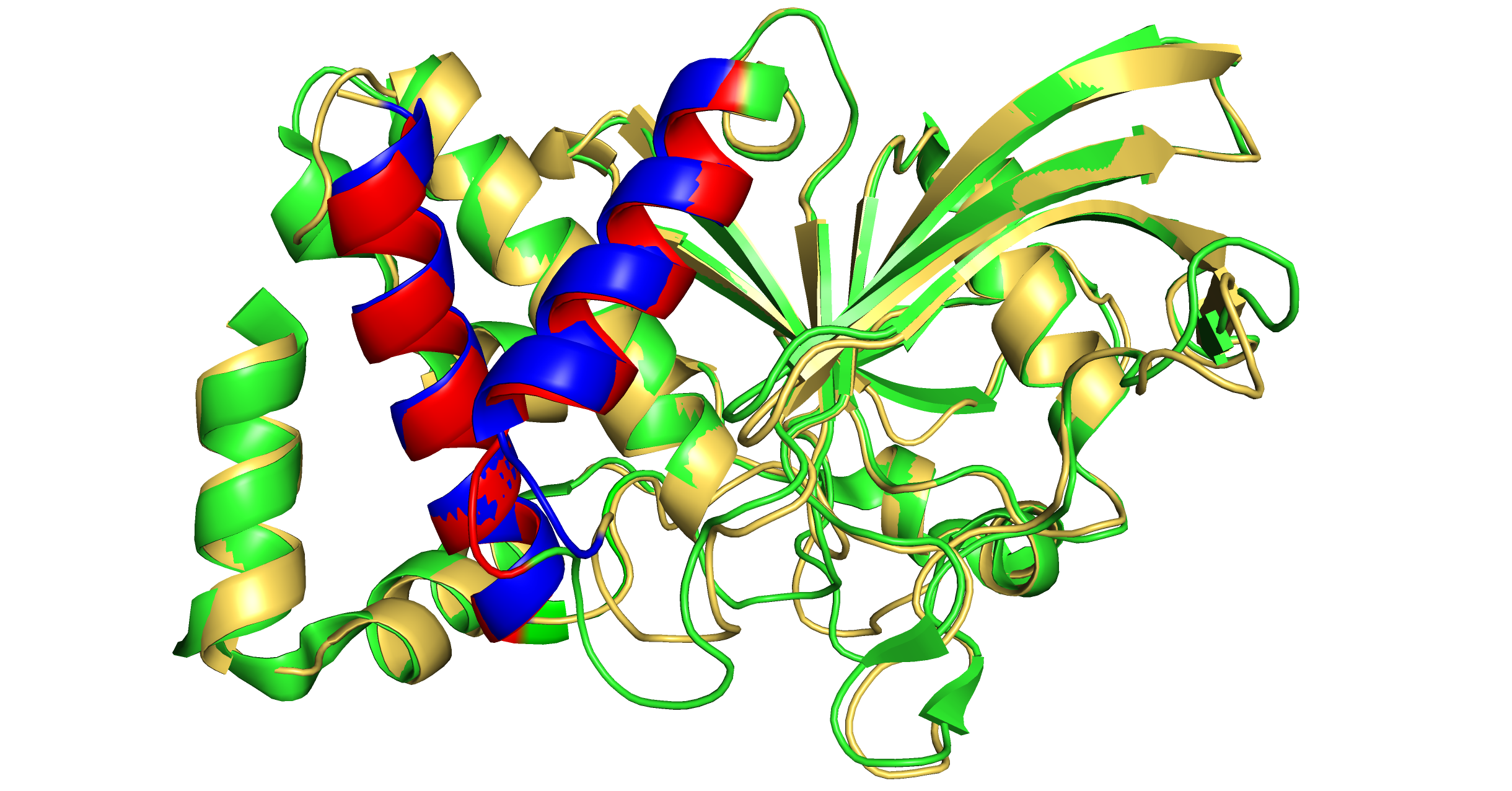}
        \caption{}
        \vspace{-3pt}
        \label{fig:nnov1}
    \end{subfigure}
    \begin{subfigure}[t]{0.49\textwidth}
        \includegraphics[width=\textwidth]{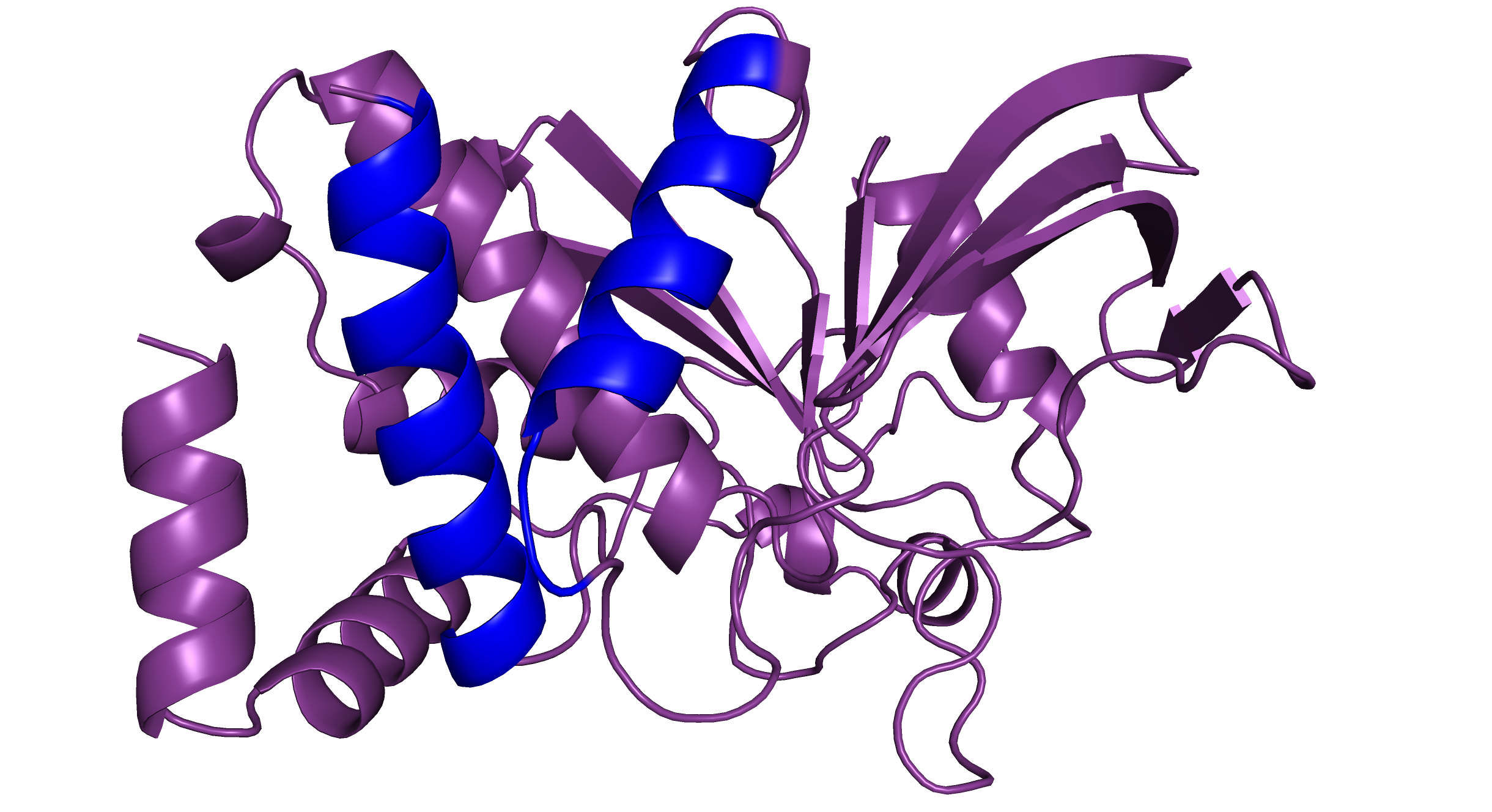}
        \caption{}
        \vspace{-3pt}
        \label{fig:nnov2}
    \end{subfigure} 

   \caption{Reconstruction of the PTP1B protein. (a) The 2nd generation of the PTP1B protein with RMSD 0.979\AA\space is shown in Table \ref{tab:alkali}. (b) The 12th generation of the PTP1B protein with RMSD 0.981\AA \space is shown in Table \ref{tab:alkali}. These generated proteins are reconstructions given the masked structural template. All of them have RMSD values below 1.0\AA. This shows that our pipeline can be used to generate context compatible protein substructures.}
  \label{fig:ptp1b_nnov}
\end{figure}

\begin{figure}[ht!] 
    \begin{subfigure}[t]{0.49\textwidth}
        \includegraphics[width=\textwidth]{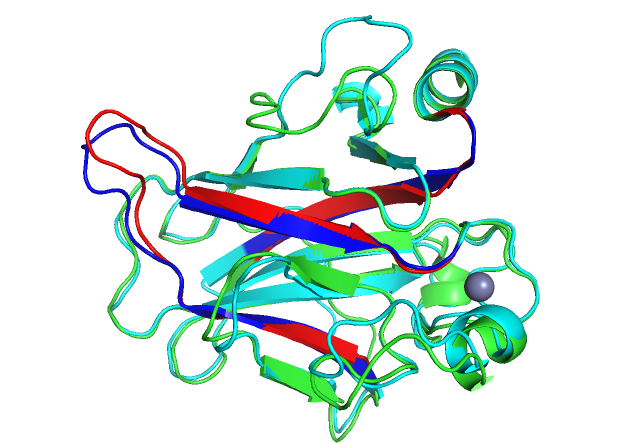}
        \caption{}
        \vspace{-3pt}
        \label{fig:phnov1}
    \end{subfigure}
    \begin{subfigure}[t]{0.49\textwidth}
        \includegraphics[width=\textwidth]{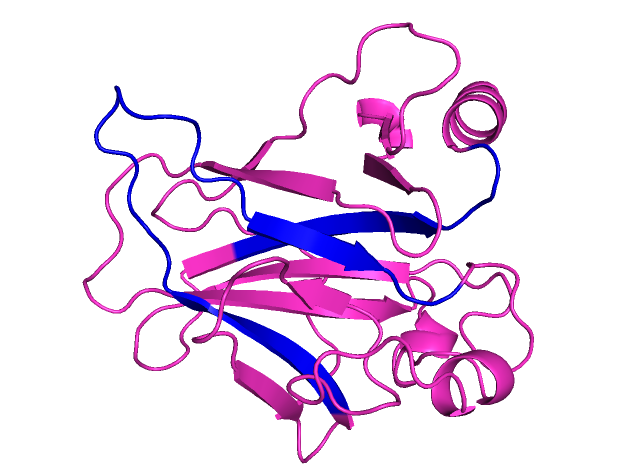}
        \caption{}
        \vspace{-3pt}
        \label{fig:phnov2}
    \end{subfigure} 

   \caption{New designed variants of P53 protein. (a) The 1st generation of the P53 protein with RMSD 3.865\AA \space is shown in Table \ref{tab:alkali}. (b) The 2nd generation of the P53 protein with RMSD 3.890\AA \space is shown in Table \ref{tab:alkali}. These proteins are highly novel because they have RMSD value far above 1.0\AA.}
  \label{fig:p53_hnov}
\end{figure}

\begin{figure}[ht!] 
    \begin{subfigure}[t]{0.49\textwidth}
        \includegraphics[width=\textwidth]{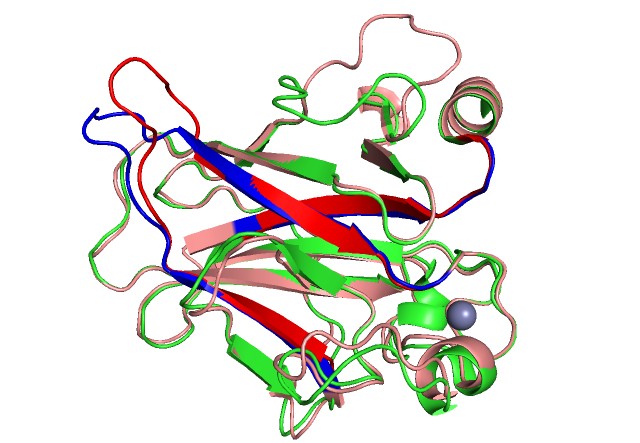}
        \caption{}
        \vspace{-3pt}
        \label{fig:pnov1}
    \end{subfigure}
    \begin{subfigure}[t]{0.49\textwidth}
        \includegraphics[width=\textwidth]{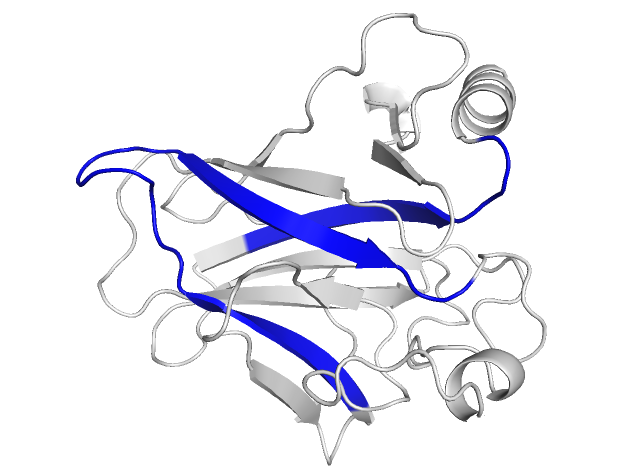}
        \caption{}
        \vspace{-3pt}
        \label{fig:pnov2}
    \end{subfigure} 
 \begin{subfigure}[t]{0.50\textwidth}
        \includegraphics[width=\textwidth]{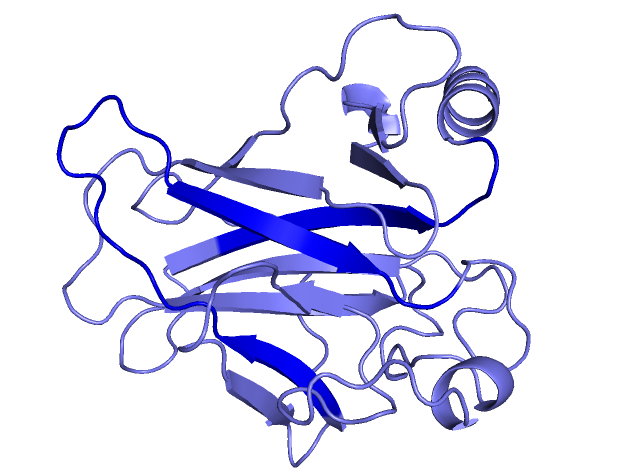}
        \caption{}
        \vspace{-3pt}
        \label{fig:pnov3}
    \end{subfigure}              
    \begin{subfigure}[t]{0.50\textwidth}
        \includegraphics[width=\textwidth]{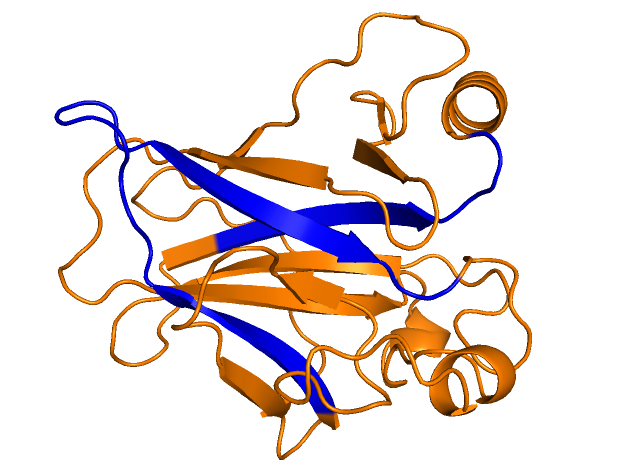}
        \caption{}
        \vspace{-3pt}
        \label{fig:pnov4}
    \end{subfigure}     

   \caption{New designed variants of the P53 protein. (a) The 3rd generation of the P53 protein with RMSD 2.384\AA\space is shown in Table \ref{tab:alkali}. (b) The 9th generation of the P53 protein with RMSD 2.089\AA\space is shown in Table \ref{tab:alkali}. (c) The 12th generation of the P53 protein with RMSD 2.001\AA\space is shown in Table \ref{tab:alkali}. (d) The 13th generation of the P53 protein with RMSD 2.115\AA \space is shown in Table \ref{tab:alkali}. These proteins are novel because they have RMSD values above 1.0\AA. }
  \label{fig:p53_nov}
\end{figure}

\begin{figure}[ht!] 
    \begin{subfigure}[t]{0.49\textwidth}
        \includegraphics[width=\textwidth]{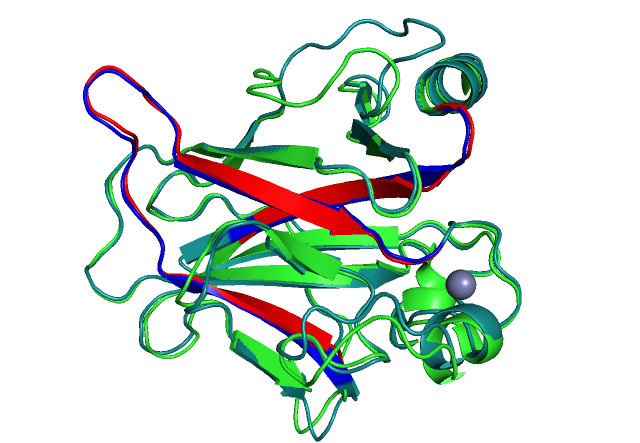}
        \caption{}
        \vspace{-3pt}
        \label{fig:pnnov1}
    \end{subfigure}
    \begin{subfigure}[t]{0.49\textwidth}
        \includegraphics[width=\textwidth]{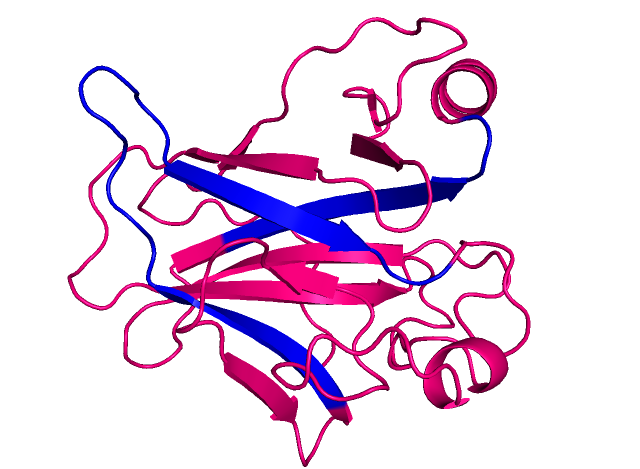}
        \caption{}
        \vspace{-3pt}
        \label{fig:pnnov2}
    \end{subfigure} 
 \begin{subfigure}[t]{0.50\textwidth}
        \includegraphics[width=\textwidth]{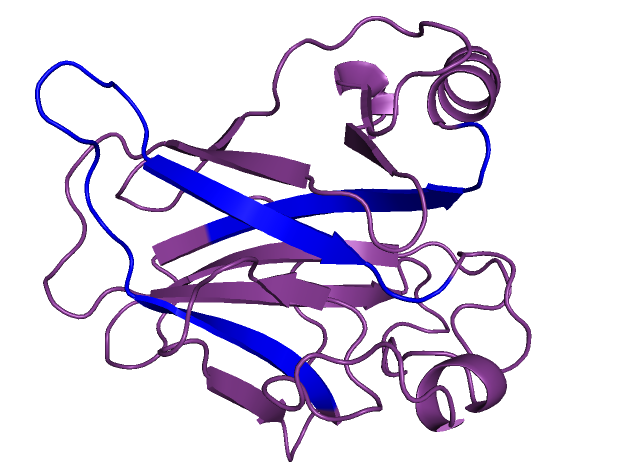}
        \caption{}
        \vspace{-3pt}
        \label{fig:pnnov3}
    \end{subfigure}              
    \begin{subfigure}[t]{0.50\textwidth}
        \includegraphics[width=\textwidth]{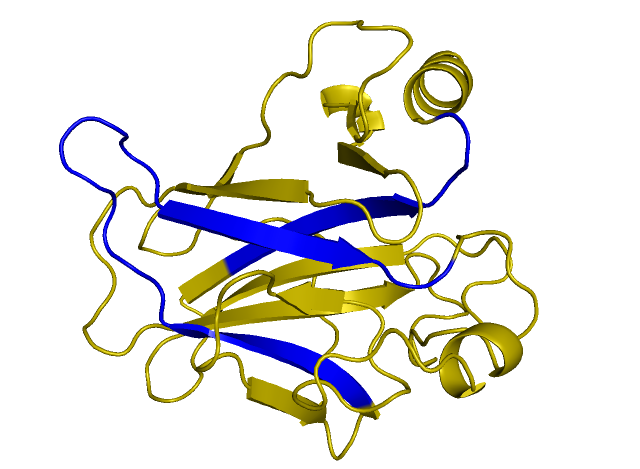}
        \caption{}
        \vspace{-3pt}
        \label{fig:pnnov4}
    \end{subfigure}     

   \caption{Reconstructions of the P53 protein. (a) The 0th generation of the P53 protein with RMSD 0.853\AA \space is shown in Table \ref{tab:alkali}. (b) The 5th generation of the P53 protein with RMSD 0.889\AA \space is shown in Table \ref{tab:alkali}. (c) The 15th generation of the P53 protein with RMSD 0.903\AA \space is shown in Table \ref{tab:alkali}. (d) The 19th generation of the P53 protein with RMSD 0.927\AA \space is shown in Table \ref{tab:alkali}. These proteins demonstrate the capability of our pipeline to identify the substrutures that are compatible to the local context. The masked local structures all have RMSD values below 1.0\AA.}
  \label{fig:p53_nnov}
\end{figure}

Although the majority of sequences are novel, we also identify some ordinary (non-novel) generations in our experiment. For the PTP1B protein generations, 2 and 12 have an RMSD lower than 1\AA, thus being non-novel protein structures. Generation 14 for PTP1B has an RMSD of 5.429\AA \space showing a highly novel protein structure. Generations 4, 6, and 17 all have RMSD values above 1 thus being novel. The P53 protein generations 0, 5, 15, and 19 have an RMSD lower than 1 which means these protein generations are non-novel protein structures. Generation 1 and 2 for the P53 protein have an RMSD of 3.865\AA \space and 3.890\AA \space respectively showing a highly novel protein structure. Generations 3, 9, 12, and 13 also have RMSD values above 1\AA \space thus being novel. Some non-novel, novel, and highly novel structures as shown in Figure \ref{fig:ptp1b_hnov}, Figure \ref{fig:ptp1b_nov}, Figure \ref{fig:ptp1b_nnov}, Figure \ref{fig:p53_hnov}, Figure \ref{fig:p53_nov}, and Figure \ref{fig:p53_nnov}).

\FloatBarrier

\subsection{Design Verification}
\begin{figure}[h] 
    \begin{subfigure}[h]{0.45\textwidth}
    \end{subfigure}
    \includegraphics[width=\textwidth]{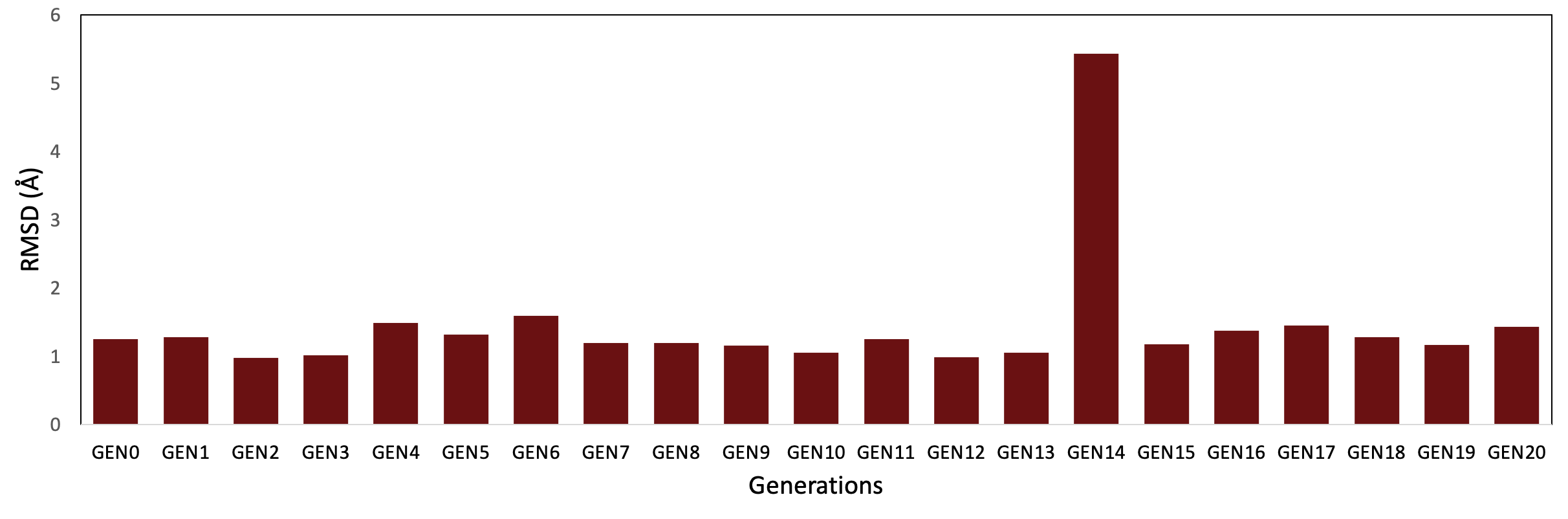}
    \caption{The RMSD value distribution of the PTP1B proteins generated in our experiments. Generation 14 has an RMSD > 3, generations 4, 6, and 17 have an RMSD > 1, and generations 2 and 12 have an RMSD < 1.}
    \vspace{-3pt}
    \label{fig:ptp1b_chart}
\end{figure}

\begin{figure}[h]
    \includegraphics[width=\textwidth]{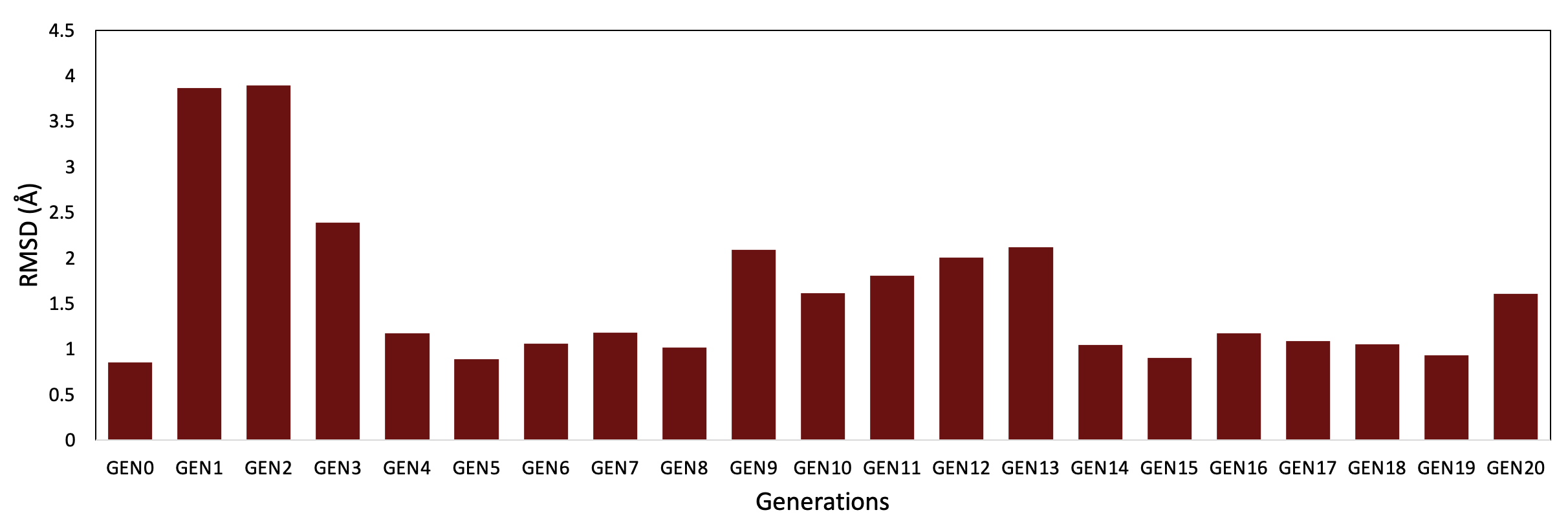}
    \caption{The RMSD value distribution of the P53 proteins generated in our experiments. Generations 1 and 2 have an RMSD > 3, generations 3, 9, 12, 13 have an RMSD > 1, and generations 0, 5, 15, and 19 have an RMSD < 1.}
    \vspace{-3pt}
    \label{fig:p53_chart}
\end{figure}

Initially, we calculated RMSD between the original structure and the generated protein structures. This approach yielded an RMSD value that was less than 1 \AA for all generated protein structures, showing that the generated structures were not novel overall. We designed a new approach where we calculated the RMSD value for only the section of the sequence that was masked and mutated using the ProteinSolver algorithm in the novel generated protein structures to the original sequences in the original protein structure. This was logical since the goal of this study was to transform the therapeutic binding pocket of the PTP1B and P53 proteins to help in drug target binding respectively. This approach yielded an RMSD value that was greater than 1 \AA for many generated protein structures, showing that the ProteinSolver model was able to generate protein structures that are novel. The distribution of our results can be seen for the PTP1B and P53 proteins respectively in Figure \ref{fig:ptp1b_chart} and Figure \ref{fig:p53_chart}.

\FloatBarrier

\subsection{Discussion}

Our protein design process is distinct from the previously mentioned studies on novel protein design \cite{cao2022design, wang2018computational, pearce2021deep, ovchinnikov2021structure, huang2022backbone} by combining the ProteinSolver, a context-aware sequence generator and AlphaFold2, a protein structure predictor, which is related but different from the approach in \cite{anishchenko2021novo}. Firstly, we attempt a potential solution for the inverse protein folding problem (from protein structure to protein sequence) whereas many studies in the current research space are focused on the translation from protein sequence to protein structure (protein folding problem). In many protein design problems, a site-related local structure is to be designed so that it can satisfy certain geometric shape for protein binding. This application scenario nicely matches our design pipeline compared to the work of \cite{anishchenko2021novo}, which focuses on generating overall new proteins.
Our design pipeline is a generative design. We generate structures that may be similar or dissimilar to the given target structure. Here the structures are predicted by AlphaFold2, which is a very slow process as it takes 1 to 1.5 hours to generate a protein structure for a given sequence. Even though our sequence generator tends to only generate geometrically feasible local structures around the masking site, it has no control how good the generated structure matches a given target structure as the sequence generator is not guided by an objective function to match the target structure. We can however, use a structure matching as implemented here to rank the generated structures based on the RMSD values and pick the candidate structures with lowest RMSD. The results in Table 1 showed that this procedure can find structures close to the target structures. Further work on this model can be done with active learning to speed up the design of the protein structure or convert the protein sequence generator into a conditional generator model. Our work can also be improved through more research on neural network architecture and hyperparameter tuning.

\section{Conclusion}
In this paper, we present a new protein design pipeline by combining ProteinSolver and AlphaFold2 to design site-related protein structures given a template structure. The ProteinSolver is used to generate context compatible sequences within the structure template while the AlphaFold2 is used to fold the generated protein sequences into protein conformations, which could be potentially used as therapeutic drug targets in diabetes and cancer. The structures generated by the deep neural network (ProteinSolver) have an average root means squared deviation (aRMSD) of 1.433\AA for the PTP1B protein and 1.606\AA for the P53 protein, two of our target structures. AlphaFold2 is used to verify that our newly designed protein sequences can fold into the desired protein structures which can be used for real-world applications. This new design pipeline can potentially help researchers and pharmaceutical scientists develop new therapeutic drugs in a time-efficient manner based on the structure of a protein target.

\section{Data and Code Availability}
All the codes can be obtained from the corresponding author with reasonable request. The network was implemented in the Python programming language using PyTorch and PyTorch Geometric libraries. The datasets and result structures can be accessed at figshare at \url{https://figshare.com/projects/Designing_novel_and_exotic_protein_structures_using_ProteinSolver_and_AlphaFold/142619}

\section{Conflict of interests}

The authors declare no conflict of interests.

\section{Author Contributions}
Conceptualization, J.H.; methodology, X.A., J.H.; software, X.A.; validation, J.H. and X.A.;  investigation, X.A., N.F.,and J.H.; resources, J.H.; data curation, X.A.; writing--original draft preparation, X.A., J.H., N.F.; writing--review and editing, J.H.; supervision, J.H.;  funding acquisition, J.H.

\section{Acknowledgement}
Research reported in this work was supported by the University of South Carolina Magellan Scholarship Grant. The views, perspective, and content do not necessarily represent the official views of the University of South Carolina.

\bibliographystyle{unsrt}  
\bibliography{references}

\end{document}